\documentclass{aa}  
\usepackage{graphicx}
\usepackage{txfonts}
\usepackage[figuresright]{rotating}
\begin{document}
   \title{Geysers in the Lagoon: new Herbig-Haro objects in M8\thanks{Based on
   observations made with the MPG/ESO 2.2-m telescope at La Silla Observatory}}

   \author{R. H. Barb\'a \inst{1}\fnmsep\thanks{Member of the Carrera del
          Investigador Cient\'{\i}fico, CONICET, Argentina}
          \and J. I. Arias \inst{1,2}\fnmsep\thanks{Fellow of CONICET, Argentina}
          }

   \offprints{R. H. Barb\'a}

   \institute{Departamento de F\'{\i}sica, Universidad de La Serena, Benavente
              980, La Serena, Chile\\
              \email{rbarba@dfuls.cl}
         \and
             Facultad de Ciencias Astron\'omicas y Geof\'{\i}sicas,
              Universidad Nacional de La Plata, and Instituto de
              Astrof\'{\i}sica La Plata, Paseo del Bosque S/N,
              B1900FWA, La Plata, Argentina\\
             \email{julia@dfuls.cl}
                          }

   \date{Received July 21, 2006; accepted June 15, 2007}

% \abstract{}{}{}{}{} 
% 5 {} token are mandatory
 
  \abstract
   {}
   {We search for direct evidence of ongoing star formation by accretion in
    the Lagoon Nebula (M8), using optical wide-field narrow-band imaging 
    obtained at La Silla Observatory.}
   {We examine [\ion{S}{ii}] and H$\alpha$ images for line-emission features
    that could be interpreted as signatures of outflow
    activity of the exciting sources.} 
   {We discover five new Herbig-Haro objects, study in detail their 
   morphology and attempt to identify their potential driving sources among 
   the population of T~Tauri stars and embedded sources in the surroundings.} 
   {The results reported here conclusively 
   demonstrate the existence of very young stars going through the accreting 
   phase in the M8 region.}


   \maketitle
%
%________________________________________________________________

\section{Introduction}
Herbig-Haro objects are direct manifestations of the 
formation process of low- and intermediate-mass stars. 
At present there is strong evidence that accretion disks surround the exciting
sources of these stellar jets (Reipurth \& Heathcote 1992, 
Reipurth \& Bally 2001, White et al. 2004, Stecklum et al. 2004, 
Kastner et al. 2005, Comer\'on \& Reipurth 2006) and that accretion energy 
powers outflows (Hartigan et al. 1994, 1995, Goodson et al. 1997, 
Turner et al. 1999). 

The vast majority of the known HH objects belong to nearby star forming 
regions (distances $\leq 500$\,pc from the Sun) and have typical sizes 
of a fraction of a parsec. 
In recent years, larger and more sensitive detectors made possible to extend 
the survey to more distant star forming regions and, more importantly,  
allowed the discovery of a new class of HH objects which stretch over 
parsec scales (Bally \& Devine 1994).

The Lagoon Nebula (Messier 8, NGC6523-NGC6530) is a prominent galactic 
\ion{H}{ii} region at a distance of 1.25\,kpc (Arias et al. 2006 and 
references therein). During the last years several studies significantly 
increased our knowledge of the young stellar content in the region. 
Specifically, Sung et al. (2000), Damiani et al. (2004), Prisinzano et al. 
(2005) and Arias et al. (2006, 2007) pointed out the presence of an abundant 
population of young stellar objects (YSOs) with typical ages of about 
$10^6$ years, distributed over the whole nebula. 
In spite of the youth suggested by many stars in M8, direct evidence of YSOs 
going through the accretion phase remained elusive. The first mention 
regarding the existence of HH objects in M8 was made by Reipurth (1981). 
More recently, Arias et al. (2006) reported the first spectroscopic 
confirmation of the HH~870 outflow, located in the Hourglass Nebula, very 
close to the ZAMS O-star Herschel~36. Additionally, Zhang et al. (2005) 
presented evidence of a molecular outflow driven by the luminous YSO M8E-IR.

In this paper we report the discovery of new optical HH outflows in M8, 
which definitely demonstrate the existence of young stars transiting the
accretion phase of their formation. The spectroscopic confirmation 
of the HH nature of some of these objects will be presented 
in a forthcoming paper.

%__________________________________________________________________

\section{Observations}

The optical images of NGC\,6530 used in this study were acquired with the 
Wide Field Imager (WFI) at the MPG/ESO 2.2-m telescope at La Silla Observatory.
The observational set consists of five [\ion{S}{ii}] and five H$\alpha$ 
exposures, each with a duration of 180 s, and five $R$ images with a exposure 
time of 10 s each. The WFI camera is an array of $2\times4$ CCD chips which 
have $2048\times4096$ 15-$\mu$m pixels each. The pixel scale is 
$0\farcs238$ pix$^{-1}$, giving a $34\times33$ square arcmin field of view. 
The images were obtained on 2000 March 17 (filters \#856, 
H$\alpha$~$\lambda6562$, and \#857, [\ion{S}{ii}]~$\lambda\lambda6717,6731$) 
under Program ID\# 2064.I-0559 and 2001 June 17 (filter $R_c$) under 
Program ID\# 165.5-0187(A). Both data sets, along with their respective bias 
and flat-field frames for calibration, were retrieved from the ESO/ST-ECF 
Science Archive Facility. The data reduction was performed using the 
IRAF\footnote{IRAF is distributed by NOAO, operated by AURA, Inc., under 
agreement with NSF} {\tt mscred} package, implemented by NOAO for specific 
processing of CCD mosaics. The images were bias subtracted and flat-fielded 
following the standard procedure provided by the {\tt mscred} user's manual 
and the WFI web page. For each filter, the images were combined into a final 
deeper mosaic free from gaps and artifacts. The FWHM of the PSFs are 1\farcs2 
and 1\farcs4 for the [\ion{S}{ii}] and H$\alpha$ frames, respectively. 
Because three of the H$\alpha$ exposures were affected by tracking 
errors, $\sim 1\%$ intensity ghost images are observed 2'' to the west 
of each object in the  final H$\alpha$ frame. These secondary images
are particularly noticeable for the brightest stars. 

%______________________________________________ Gamma_1 (lg rho, lg e)
%______________________________________________ Figura 1 (mosaico de M8)
   \begin{sidewaysfigure*}
   \centering
   \includegraphics[width=23cm]{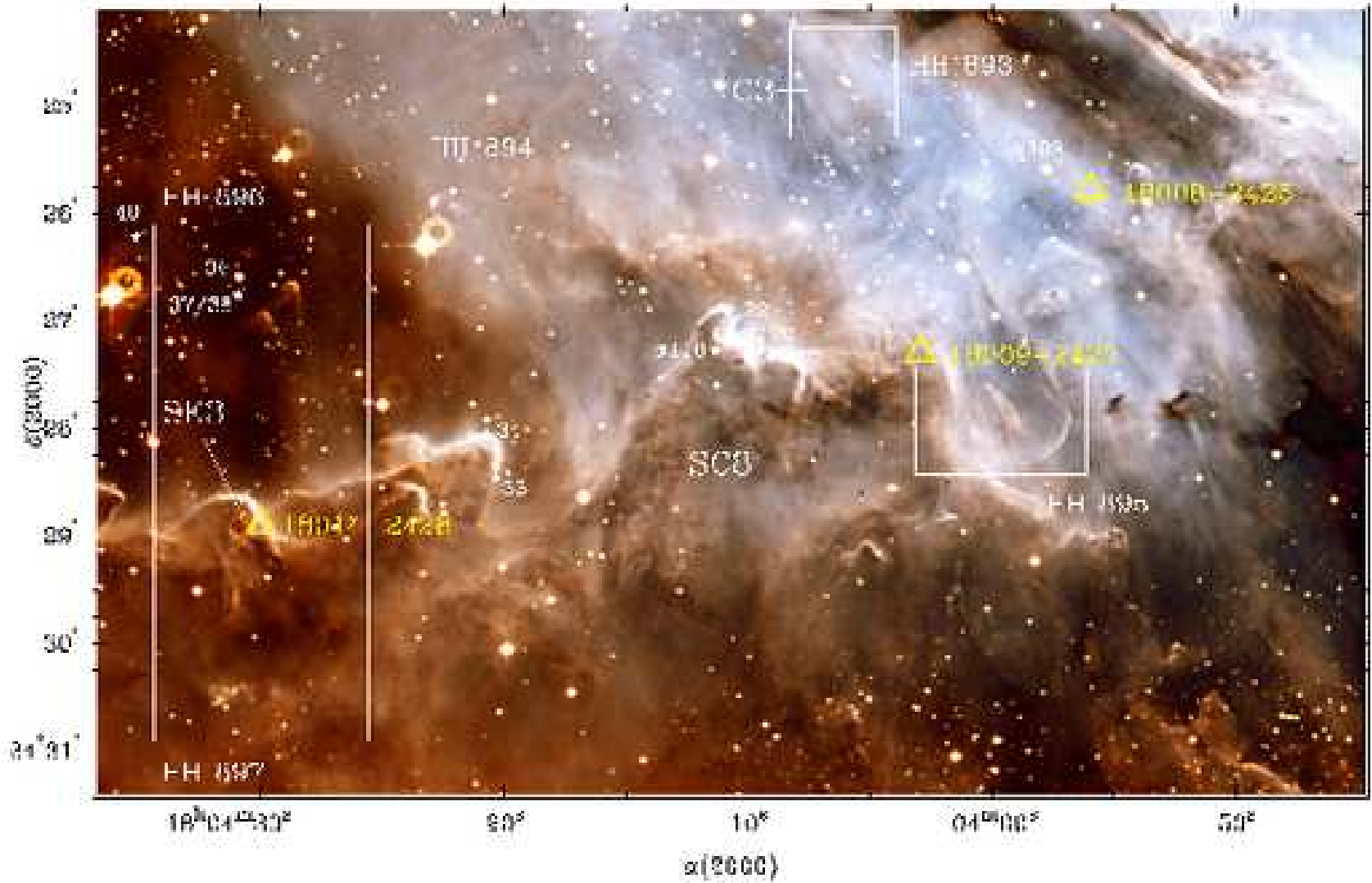}
   \caption{Three-colour image of the M8 nebula taken with the WFI at the 
MPG/ESO 2.2-m telescope, showing [\ion{S}{ii}] emission ({\em red}), 
H$\alpha$ emission ({\em blue}), and  [\ion{S}{ii}]+H$\alpha$ emission 
({\em green}). The rectangles indicate the regions where the new HH~objects 
are found. Some T~Tauri stars identified in the area are denoted with ``star'' 
symbols and labelled according to the numbers from Arias et al. 2007. 
The triangles represent the IRAS sources of interest in this study. 
The position of the molecular clumps C3, SC8 and SE3 (Tothill et al. 2002) 
are also marked.
The ``donuts''  next to the bright stars are artifacts due to internal 
reflections in the filters.}
              \label{s2}%
    \end{sidewaysfigure*}

The images were not flux calibrated. In order to compare the  
[\ion{S}{ii}] and H$\alpha$ images, we constructed the difference map
between H$\alpha$ and four times the [\ion{S}{ii}] image, which gives a good
picture of how both emission lines are related.
 
Celestial coordinates for the images were obtained from the 2MASS Point Source
Catalogue (Skrutskie et al. 2006), using the {\tt msccmatch} task. 
Only objects with $K_s<12$ and photometric quality flag {\em ``AAA''} were 
considered. In order to estimate the astrometric accuracy, we positionally 
matched the stars in our catalogue with those in both the 2MASS PSC and the 
Guide Star Catalogue Version 2.2.01 (GSC 2.2, STScI, 2001). 
The rms residuals between the positional tables from this work and from the
former  data bases are found to be $\sim 0\farcs4$ in both coordinates.

%------------------------------------------------------------------------------
\section{Identification of new Herbig-Haro objects}

Figure~\ref{s2} shows a three-colour image of the M8 nebula, with 
H$\alpha$ in blue, [\ion{S}{ii}]~$\lambda\lambda6717,6731$ in red, and 
[\ion{S}{ii}]+H$\alpha$ in  green. 
The image exhibits an extremely rich and complex structure.
The visual inspection of the individual narrow-band images allowed us to 
distinguish several nebular features, such as ionization fronts, ``fingers'', 
dark and bright globules, etc. Some of them could be identified as HH objects.
 
\begin{figure*} 
   \centering
   \includegraphics[width=13cm]{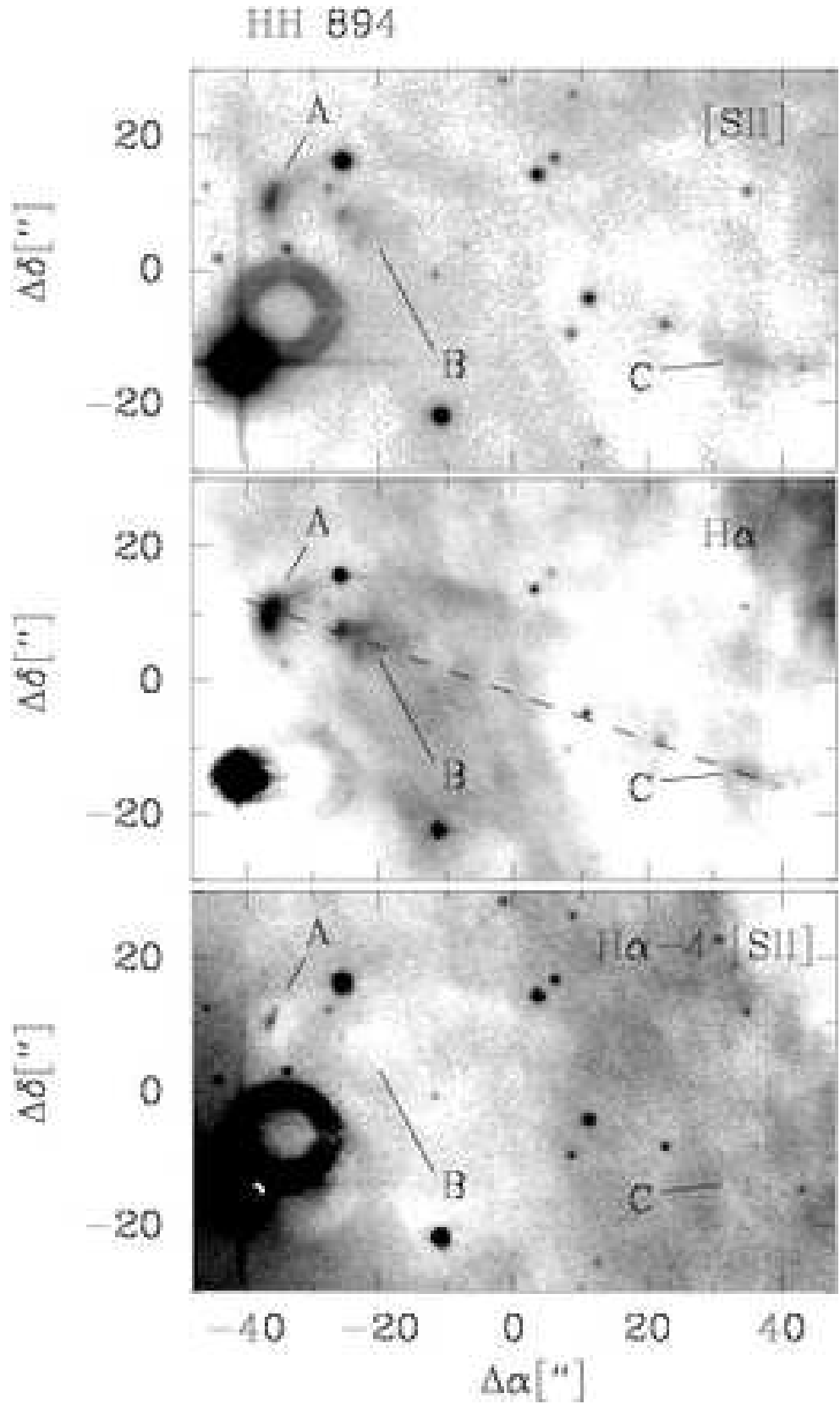}
   \caption{Three close-ups of HH~894. {\em Top and middle:} [\ion{S}{ii}] and 
   H$\alpha$ images, respectively. In these panels, light colours indicate 
   low emission values and dark colours correspond to high emission values. 
   The labels A, B and C, refer to the nebular components described in the 
   text. In the H$\alpha$ image, a dashed line joining these components has 
   been marked. {\em Bottom:} Difference image obtained by subtracting 4 times 
   the [\ion{S}{ii}] image from the H$\alpha$ image. Light colours represent 
   features dominated by H$\alpha$ emission, and dark colours correspond to 
   features dominated by [\ion{S}{ii}] emission. Stars appear black because 
   the [\ion{S}{ii}] image was multiplied by a factor of 4 before subtraction.
   In the three panels, the ``donut''  next to the bright star on the left is 
   an artifact due to internal reflections in the filters.}
              \label{hh2}%
    \end{figure*}

\begin{figure*} 
   \centering
   \includegraphics[width=13cm]{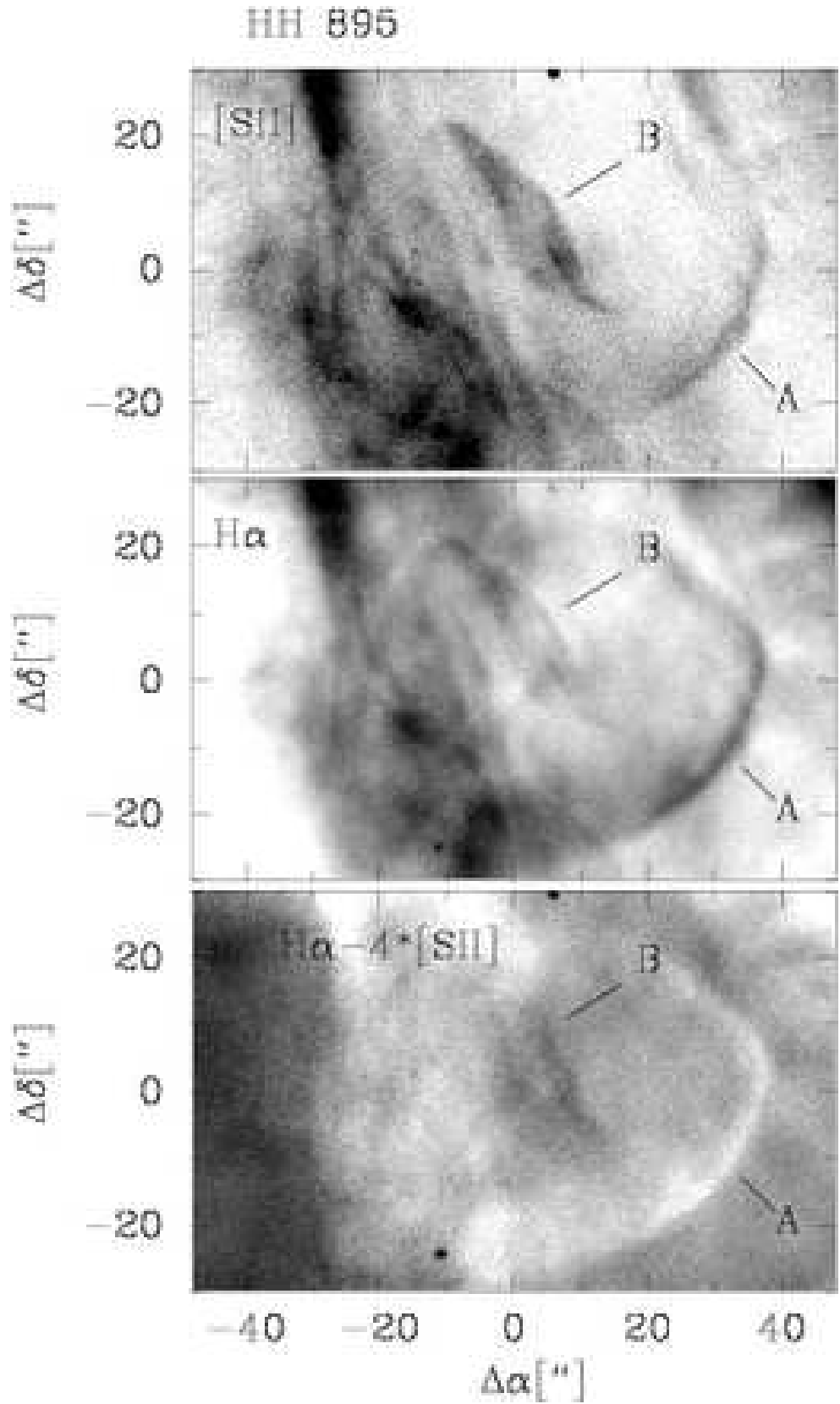}
   \caption{Three close-ups of HH~895. {\em Top and middle:} [\ion{S}{ii}] and 
   H$\alpha$ images, respectively. In these panels, light colours indicate 
   low emission values and dark colours correspond to high emission values. 
   The labels A and B, refer to the nebular components described in the 
   text.  {\em Bottom:} Difference image obtained by subtracting 4 times 
   the [\ion{S}{ii}] image from the H$\alpha$ image. Light colours represent 
   features dominated by H$\alpha$ emission, and dark colours correspond to 
   features dominated by [\ion{S}{ii}] emission. Stars appear black because 
   the [\ion{S}{ii}] image was multiplied by a factor of 4 before subtraction.
}
              \label{hh3}%
    \end{figure*}

\begin{figure*}
   \centering
   \includegraphics[width=17cm]{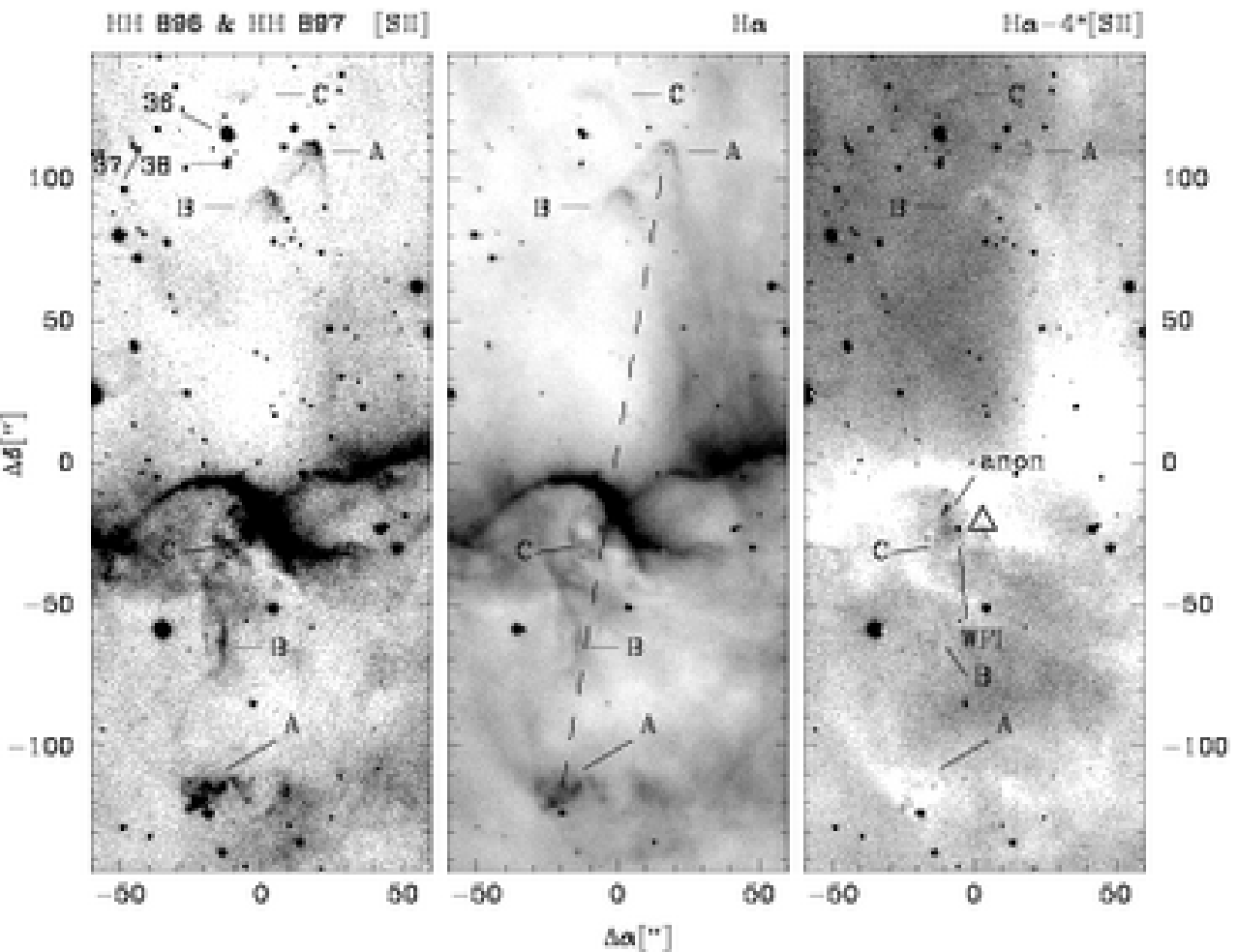}
   \caption{Three images of the HH~896 and HH~897 outflows. The northern 
   nebular features correspond to HH~896, whereas the southern ones form 
   HH~897. For each case, the labels A, B and C, refer to the subcomponents 
   described in the text. {\em Left:} [\ion{S}{ii}] image. Light colours 
   represent low emission values and dark colours correspond to high emission 
   values. The numbers indicate some T~Tauri stars identified by Arias et
   al. (2007). {\em Middle:} H$\alpha$ image. As in the previous panel, light 
   and dark colours correspond to low and high emission values, respectively. 
   The dashed line represents the presumed axis of the bipolar jet that both
   HH objects could be forming  
   (see Sec.~\ref{parsec-scale}). {\em Right:} Difference image obtained by 
   subtracting 4 times the [\ion{S}{ii}] image from the H$\alpha$ image. Light 
   colours represent features dominated by H$\alpha$ emission, and dark 
   colours correspond to features dominated by [\ion{S}{ii}] emission. Stars 
   appear black because the [\ion{S}{ii}] image was multiplied by a factor of 
   4 before subtraction. ``WFI'' and ``anon'' refer to the two stellar objects
   found in the $850\,\mu$m clump SE3 (see text). The triangle marks the
   position of the IRAS source  $18014-2428$.}
              \label{hh4}%
    \end{figure*}

The newly identified  HH objects are primarily located in the southern edge of 
M8, formed by a group of bright-rimmed clouds defined as ``Southeastern 
Bright Rim'' and ``Extended Bright Rim'' by Lada et al. (1976).
Tothill et al. (2002) resolved several continuum (850\,$\mu$m) and CO 
clumps in this area. The presence of several classical and weak 
T~Tauri stars and Herbig~Ae/Be objects was also reported 
(Arias et al. 2007, Sung et al. 2000 and references therein).  
The rectangles in Figure~\ref{s2} indicate the regions where the new HH
flows are found. These objects are shown in detail in Figures~\ref{hh1} to
\ref{closeup}. Their equatorial coordinates and some related notes
are given in Table~\ref{coordHH}.

In this section we describe the morphological properties of the new 
HH objects and present some speculation about their potential driving sources.

\subsection{HH~893}

\begin{figure*}  
   \centering
   \includegraphics[width=17cm]{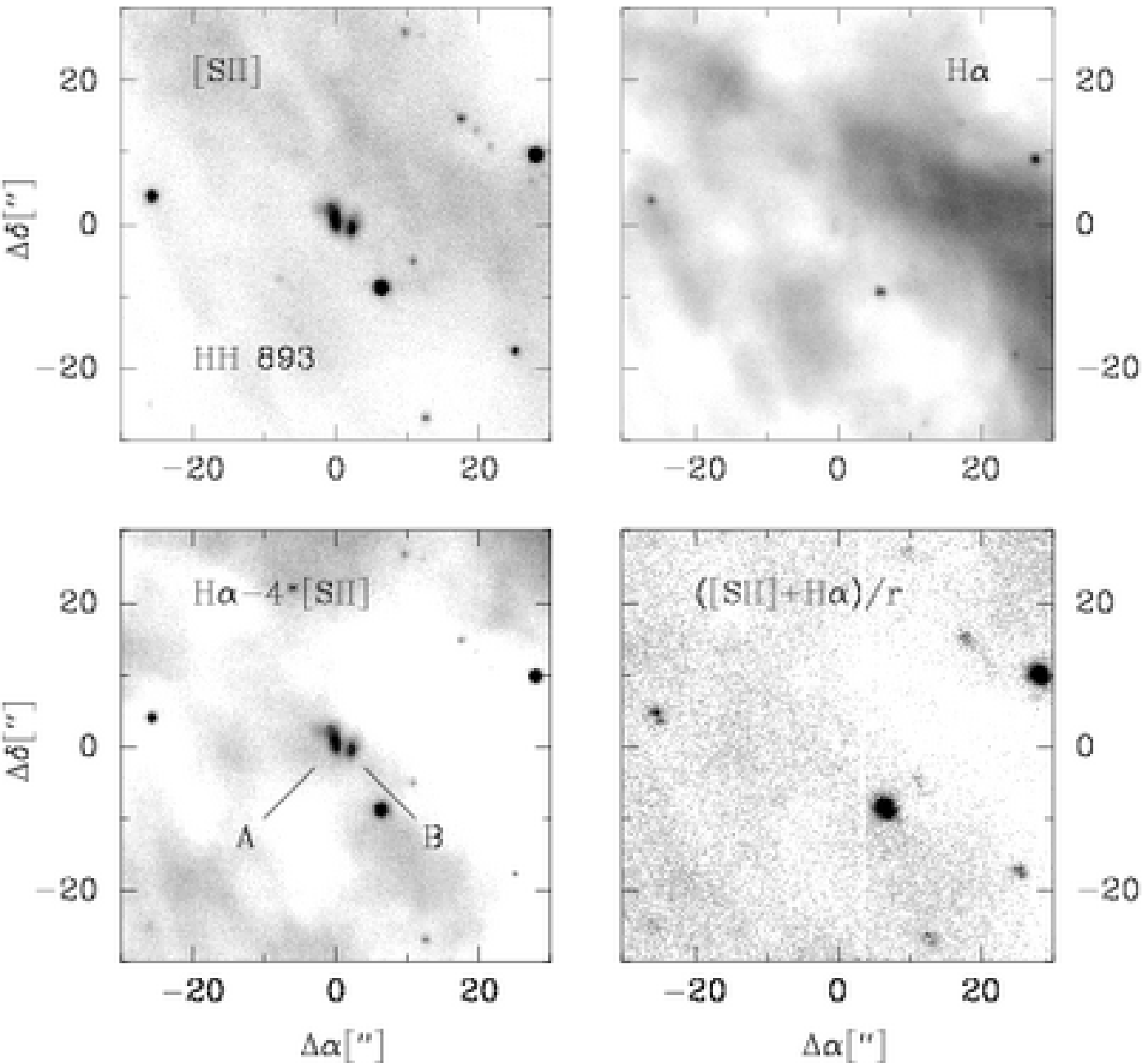}
   \caption{Four close-ups of HH~893.  {\em Top:}  [\ion{S}{ii}] (left) and 
   H$\alpha$ (right) images. Light and dark colours indicate low and high 
   emission values, respectively. {\em Bottom left:} Difference image obtained 
   by subtracting 4 times the [\ion{S}{ii}] image from the H$\alpha$ image. 
   Light colours represent features dominated by H$\alpha$ emission,
   and dark colours correspond to features dominated by [\ion{S}{ii}]
   emission. Stars appear black because the [\ion{S}{ii}] image was multiplied
   by a factor of 4 before subtraction. The labels A and B, refer to the 
   components described in the text. {\em Bottom right:} Ratio image 
   between the sum [\ion{S}{ii}] + H$\alpha$ and the $R$-band image. In this 
   panel, light colours correspond to features dominated by line emission, 
   while dark colours mean that a continuum light source is present. 
   }
              \label{hh1}%
    \end{figure*}

HH~893 is a small object situated close to a set of bright H$\alpha$ 
filaments. Detailed views of this object are shown in Figure~\ref{hh1}.
HH~893 is barely detectable in the H$\alpha$ and $R$ images.  
In the [\ion{S}{ii}] image, it is composed of two bright emission knots 
(A and B) separated by $\sim 1"$.
Its size is approximately $6''\times4''$ ($7500\times5000$\,AU at the 
distance of M8), the eastern lobe A being larger. 

In order to investigate the emission characteristics of this 
object, we considered the ratio of H$\alpha$ + [\ion{S}{ii}] to the $R$-band
flux (bottom right panel of Figure~\ref{hh1}). 
If there were significant scattered light (from any existing continuum source),
this ratio would depart from unity. However, the observed ratio is around 
unity, implying that the filaments that form HH~893 are entirely emission 
knots.

We could not identify any convincing driving source for HH~893. 
As the M8 nebula is seen projected against the galactic bulge, 
a dense population of background infrared sources is observed. 
The nearest 2MASS source is located $2\farcs2$ to the west of the
feature $B$ of HH~893, but its near-IR colours correspond to a typical 
reddened background giant ($J-H=2.04$, $H-K_s=0.83$). 
We note here that HH~893 is located $\sim20''$ to the west of the 
$850\,\mu$ clump C3 (Tothill et al. 2002) and that the nearest IRAS source
($18008-2425$) is about $2\farcm7$ distant (see Figure~\ref{s2}).

%______________________________________________ Figura 2 (HH 893)
%

%______________________________________________ Figura 3 (HH 894)
%

%______________________________________________ Figura 4 (HH 895)
%_______________________________________________ Figura 5 (HH 896 y HH 897)

\subsection{HH~894}

HH~894 shows remarkable characteristics. In Figure~\ref{hh2} we present 
three images showing different aspects of this object. 
HH~894 consists of various components, which are labeled A, B and C in this 
figure. The structures A and B consist of a set of filaments that curve to form
bow-shaped features brightest at their tips. 
While feature A shows two extended parabolic tails, feature B appears much
more diffusely. The component B may be in turn decomposed in three smaller
knots.  The working surfaces are more evident in the H$\alpha$ image.
Tracking back about $68''$ along the line joining the bright tips of the 
knots A and B, a small [\ion{S}{ii}] linear feature arises. 
This 10'' size knot is labeled C in Figure~\ref{hh2}. 
If features A, B, and C were part of the same outflow, 
then the projected size of this outflow would be of approximately 80'', 
which corresponds to $\sim0.5$\,pc at the distance of M8. 

Several faint optical and near-IR stars are found lying on the presumed axis
of this jet. This presumed axis also intersects 
the tip of a dusty pillar structure known as clump SC8 (Tothill et al. 2002), 
where three emission line stars, ABM\,20, ABM\,22 (Arias et al. 2007) and 
LkH$\alpha$\,110 (Herbig 1957), are located (see Figure~\ref{s2}).
ABM\,20 and ABM\,22 were recently identified as T~Tauri stars. The latter is 
additionally a very particular object since it shows asymmetric 
forbidden emission lines, which might be a signature of HH outflows. 
This naturally leads to speculation that HH~894 is driven by some of the 
sources located at the tip of SC8, in which case its projected 
length would be of about 3' or 1.1\,pc at the distance of M8. 
Future spectroscopic studies will help to answer this question.

\subsection{HH~895}

HH~895 shows a very peculiar morphology. In Figure~\ref{hh3} we present three 
detailed views of this intriguing object.
It consists of a large bow-shaped arc (A) with a central condensation (B).
The arc, whose amplitude is of approximately $50''$, is dominated by H$\alpha$ 
emission and  hence appears whitish in the H$\alpha$~-~4*[\ion{S}{ii}]
difference image shown in the bottom panel of Figure~\ref{hh3}.
The component B is curious rhomboidal structure, whose western edge 
looks brighter in [\ion{S}{ii}].
We note here that HH~895 is found in a region where a complex network of 
filaments exists. A further analysis is needed to disentangle the several 
components present in this area.

\subsection{HH~896}

Two nebular features are found roughly equally distant north and south 
from the $850\,\mu$ clump SE3 identified by Tothill et al. (2002). The IRAS 
source $18014-2428$ is also associated with this molecular clump (see 
Figures~\ref{s2} and \ref{hh4}). The northern feature, located about
$2\farcm2$ from IRAS $18014-2428$, is the HH~896 outflow. 

\begin{figure*}  
   \centering
   \includegraphics[angle=-90, width=17cm]{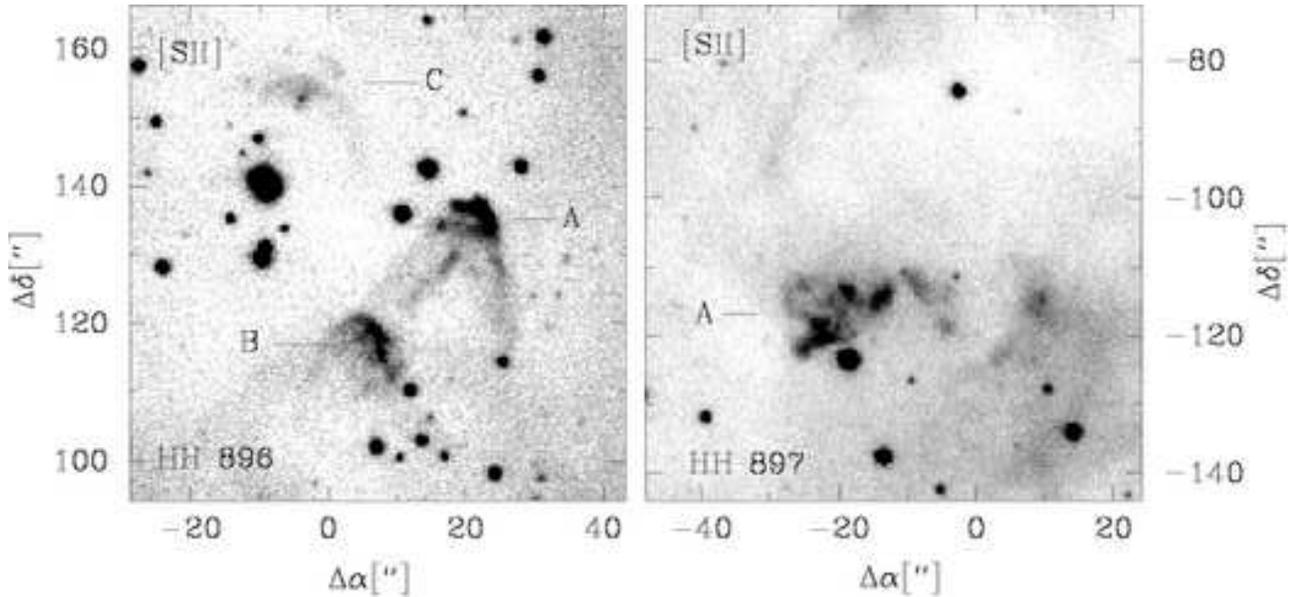}
   \caption{ Close-ups of HH~896 ({\em left}) and HH~897 ({\em right}) in the 
   light of [\ion{S}{ii}]. Light and dark colours indicate low and high 
   emission values, respectively. The letters refer to the 
   subcomponents described in the text.}
              \label{closeup}%
    \end{figure*}

The right panel of Figure~\ref{closeup} presents a close-up of HH~896 in the 
light of [\ion{S}{ii}]. This object consists of two well-defined parabolic bow 
shocks (A and B), along with a fainter nebular arc (C).
The latter is about $5''$ north from the stars ABM\,36, ABM\,37 and ABM\,38,
which have been recently classified as T~Tauri stars by Arias et al. (2007).
The morphologies of the HH~896 A and B components clearly resemble those 
observed in the HH\,1/2-NW and HH\,214W objects (Ogura 1995). 
They are appreciably bright both in H$\alpha$ and 
[\ion{S}{ii}] and present a knotty structure, 
In addition, their axis of symmetry point 
toward IRAS $18014-2428$, although there is a slight difference in 
their apparent directions.
  
Perhaps it is premature to give an interpretation regarding the  nature
of the multiple bow shocks in HH 896, but the 
difference in the apparent axis of symmetry of the components A and B 
could be related to the phenomenon of jet bending due to
the motion of the source within a binary system.
This mechanism was proposed by Fendt \& Zinnecker (1998) to explain 
the misalignment between the apparent direction of propagation 
for jet and counter jet. 
Anyway we cannot affirm, based on imaging alone, that the bow shocks A and B
are produced by the two components of a binary system associated with
IRAS $18014-2428$.
Further spectroscopic studies are needed to determine a convincing  
association between HH 896 and its actual exciting source.

%_______________________________________________ Figura 6 
%_______________________________________________

\subsection{HH~897}

HH~897 is situated to the south of the molecular clump SE3 and the IRAS 
source $18014-2428$  (see Figures~\ref{s2} and \ref{hh4}). 
Three nebular emission features (A, B and C) can be identified as subcomponents
of HH~897. 

Located about $1\farcm6$ south of IRAS $18014-2428$, the component A have an
irregular knotty morphology. A close-up of this intriguing feature is shown 
in Figure~\ref{closeup}.
There is a pair of nebular structures (B and C) connecting the component A 
with the mentioned IRAS source.  
The feature B consist of two parallel filaments, about $14''$ long, the 
western one being somewhat larger. 
The feature C is directly associated with the area of IRAS $18014-2428$ and 
the X-ray source WFI\,11091 (Prisinzano et al. 2005; Damiani et al. 2004, 
source 609). 
It is a rather complex structure that apparently consist of two faint arcs 
along with several smaller knots. 
Finally, it is worthwhile to mention that there is a faint ``anonymous'' 
stellar object, located $8\farcs5$ northeast from WFI\,11091, 
from which a $\sim3''$ long filament seems to develop (P.A.=$150\deg$).

\subsection{A parsec-scale outflow?}
\label{parsec-scale}

As described in the previous sections, HH~896 and HH~897 are more or less 
symmetrically located from the molecular clump SE3 (Tothill et al. 2002)
and the IRAS source $18014-2428$. It seems fairly likely, based on the 
observed morphology, that both objects constitute the jet and the counter jet
of a single bipolar flow\footnote{Spectroscopy recently obtained with 
Magellan-IMACS (LCO) confirm this conjecture and will be presented in a 
forthcoming paper.}. 
In Figure~\ref{hh4}, the presumed axis of this bipolar jet has been
approximately indicated as a dashed line.
IRAS $18014-2428$ appears as a potential driving source for this outflow. 
Under this speculation, projected on the sky, the jet formed by HH~896 and
HH~897 would have a total length of about $3\farcm9$ or roughly 1.4 pc at the 
distance of M8, being an example of parsec-scale outflow.

\section{Summary and conclusions}

We report the discovery of five new Herbig-Haro objects in the 
M8 nebula at 1.25~kpc.
These objects, for which the numbers HH~893, HH~894, HH~895, HH~896
and HH~897 have been assigned in the HH~catalogue, show the
following characteristics.

{\em HH~893.} --  This is a small feature composed of two nebular emission 
knots. Its connection with nearby T~Tauri stars and/or IRAS sources is rather 
difficult to discern. 

{\em HH~894.} -- This is a peculiar object which consist of three 
approximately aligned nebular emission features. The presumed axis of this 
jet intersects the tip of the molecular clump SC8 (Tothill et al. 2002), where 
the T~Tauri stars ABM\,20 and ABM\,22 (Arias et al. 2007) are located. This 
leads to the speculation that HH~894 is driven by some of the former young 
stars, although spectroscopic studies are  obviously needed. 

{\em HH~895.} -- This is an intriguing object with a very peculiar morphology,
located in a region full of filaments and other nebular features.

{\em HH~896.} -- This object consists of three bow-shaped arcs. The faintest
one (C) is $\sim 5"$ north from a group of recently discovered T~Tauri stars. 
The other two features (A and B) clearly resemble larger bow shocks commonly
seen in HH objects. Their axis of symmetry point toward the IRAS source 
$18014-2428$, located in the molecular clump SE3 (Tothill et al. 2002).

{\em HH~897.} -- This object consist of three emission features with irregular
and knotty morphologies. It might be associated with IRAS $18014-2428$, 
as well as with the X-ray source WFI\,11091 (Damiani et al. 2004) and another
unidentified stellar object present in the clump SE3. 

HH~896 and HH~897 are more or less symmetrically located from the molecular 
clump SE3 and the IRAS source $18014-2428$. 
Based on the observed morphology, it is proposed that these objects constitute 
the jet and the counter jet of a single parsec-scale bipolar outflow, whose 
projected total length would be of about $3\farcm9$ or roughly 1.4 pc at the 
distance of M8. In a forthcoming paper, we will present spectroscopy obtained
with Magellan-IMACS (LCO) that confirms this hypothesis.

The IRAS source $18014-2428$ appears as a potential driving source for this
parsec-scale jet. Molinari et al. (1996) reported an ammonia maser 
associated with IRAS $18014-2428$. They also classified it as a ``bonafide'' 
protostar and derived a FIR luminosity of $1.7\,10^4\,L_\odot$. 
However the kinematic distance of 2.87\,kpc determined for this source might 
be overestimated, which would imply an overestimation of its luminosity too. 

The large projected distances of the objects HH~896 and HH~897
from their potential exciting sources lead to very large dynamical ages.
Assuming for the flow speed of the jets a ``standard'' propagation velocity 
of 300 km\,s$^{-1}$, we derive a dynamical age of $4.6 \times 10^3$ 
years for the HH~896/HH~897 system.
A future kinematic study through radial velocities and 
proper motions analysis will certainly help to constrain the numbers 
suggested here. 

The discovery of HH objects in M8 conclusively demonstrate the existence 
of very young stars going through the accretion phase of their
formation. Finally, we can assert that the 
presence of large-scale outflows makes the M8 nebula an especially attractive 
target for the study of this and other sorts of stellar formation activity.

%__________________________________________________ One column table
   \begin{table}
      \caption[]{New Herbig-Haro objects in M8.}
         \label{coordHH}
     $$ 
         \begin{array}{p{0.13\linewidth}llp{0.4\linewidth}}
            \hline
            \noalign{\smallskip}
            Object   &  \alpha_{2000}  & \delta_{2000} & Notes\\
            \noalign{\smallskip}
            \hline
            \noalign{\smallskip}
            HH~893 & & & \\
            ~~~~A  & 18\,04\,06.12 & -24\,24\,46 & {\rm [SII] knot}  \\
            ~~~~B  & 18\,04\,05.95 & -24\,24\,47 & {\rm [SII] knot}  \\
            \noalign{\smallskip}
            HH~894  & & &  \\
            ~~~~A  & 18\,04\,22.87 & -24\,25\,52 & {\rm parabolic bow-shock} \\
            ~~~~B  & 18\,04\,22.01 & -24\,25\,55 & {\rm knots}   \\ 
            ~~~~C  & 18\,04\,17.69 & -24\,26\,16 & {\rm filament}   \\
            \noalign{\smallskip}
            HH~895 & & & \\
            ~~~~A  & 18\,03\,57.21 & -24\,28\,04 &  {\rm wide bow-shock} \\
            ~~~~B  & 18\,03\,59.16 & -24\,27\,53 &  {\rm knotty filament} \\
            \noalign{\smallskip}
            HH~896 & & & \\
            ~~~~A  & 18\,04\,28.60 & -24\,26\,38 & {\rm parabolic bow-shock} \\
            ~~~~B  & 18\,04\,29.75 & -24\,26\,57 & {\rm parabolic bow-shock} \\
            ~~~~C  & 18\,04\,30.40 & -24\,26\,20 & {\rm faint bow-shock} \\
            \noalign{\smallskip}
            HH~897 & & & \\
            ~~~~A  & 18\,04\,31.40 & -24\,30\,27 & {\rm knotty bow-shock}\\
            ~~~~B  & 18\,04\,31.00 & -24\,29\,33 & {\rm filaments}\\
            ~~~~C  & 18\,04\,30.90 & -24\,28\,59 & {\rm arcs and knots}\\
          \noalign{\smallskip}
            \hline
         \end{array}
     $$ 
%\begin{list}{}{}
%\item[$^{\mathrm{a}}$] This is footnote a
%\end{list}
   \end{table}

\begin{acknowledgements}
We thank the anonymous referee for many comments and suggestions that have 
improved this paper. 
This publication makes use of data products from the Two Micron All Sky
Survey, which is a joint project of the University of Massachusetts and
the Infrared Processing and Analysis Center/California Institute of
Technology, funded by the National Aeronautics and Space Administration
and the National Science Foundation. This research has also made use of
Aladin and Simbad Database, operated at CDS, Strasbourg, France.
Financial support from FONDECYT No. 1050052 and from PIP-CONICET No. 5697
is acknowledged by RHB and JIA, respectively. 
This paper was written during 
the 10th Workshop of the International Program of Advanced Astrophysics
``Guillermo Haro'' (INAOE, Tonantzintla, M\'exico). 
The authors are grateful to the organizing committee for the warm and friendly 
hospitality.
\end{acknowledgements}

\end{document}